\documentclass[aps,twocolumn,pre,floatfix,superscriptaddress]{revtex4-1}

\usepackage{graphicx}

\usepackage{amsmath,amsfonts,amssymb}

\newcommand{\bra}[1]{\langle\left.{#1}\right|}
\newcommand{\ket}[1]{\left|{#1}\right.\rangle}

\begin{document}

\title{Eigenstate Thermalization Scaling in Majorana Clusters: from Chaotic to Integrable Sachdev-Ye-Kitaev Models}


\author{Masudul Haque}

\affiliation{Max Planck Institute for the Physics of Complex Systems, N\"{o}thnitzer Str. 38, 01187 Dresden}

\affiliation{Department of Theoretical Physics, Maynooth University, Co. Kildare, Ireland}

\author{P.~A.\ McClarty}

\affiliation{Max Planck Institute for the Physics of Complex Systems, N\"{o}thnitzer Str. 38, 01187 Dresden}


\begin{abstract}

  The eigenstate thermalization hypothesis (ETH) is a conjecture on the nature of isolated quantum
  systems that guarantees the thermal behavior of subsystems when it is satisfied. ETH has been
  tested in various forms on a number of local many-body interacting systems. Here we examine the
  validity of ETH in a class of nonlocal disordered many-body interacting systems --- the
  Sachdev-Ye-Kitaev Majorana (SYK) models --- that may be tuned from chaotic behavior to
  integrability. Our analysis shows that SYK$_4$ (with quartic couplings), which is maximally
  chaotic in the large system size limit, satisfies the standard ETH scaling while SYK$_2$ (with
  quadratic couplings), which is integrable, does not.  We show that the low-energy and high-energy
  properties are affected drastically differently when the two Hamiltonians are mixed.

\end{abstract}

\maketitle

\section{Introduction}
\label{sec:introduction}

For well over a century it has been understood that physical observables of a system coupled to a
heat bath in the long time limit are such that they may be computed from a statistical average over
all states of the system with a weight that depends on only a few parameters including the
temperature of the bath.

Since heat baths are idealized objects coupled to but distinct from the system of interest, a natural question is: under what circumstances can a subsystem of a closed (isolated) quantum system be
thermalized by the remainder, in the sense of physical observables being expressed in terms of a
trace over some density matrix with a handful of constraints including particle number and
temperature? This led to the eigenstate thermalization hypothesis (ETH) which is a conjecture about
the nature of matrix elements of physical observables that, if true, reconciles the predictions of
thermodynamics (with respect to an ensemble whether microcanonical, canonical or grand canonical)
with those of quantum mechanics in the long-time limit \cite{PhysRevE.50.888,
  PhysRevA.43.2046, d2016quantum}.

The essence of ETH is that diagonal matrix elements of physical observables in the energy
eigenstate basis, that is distinguished by time evolution, are expected to be quasi-randomly
distributed with a width that falls off rapidly with the Hilbert space dimension leading to a smooth
variation of the observable with energy. The off-diagonal matrix elements are also expected to fall
off rapidly with increasing system size which can be viewed as a de-phasing of the subsystem by the
remainder of the system.  Taken together, these conditions on the matrix elements ensure that
subsystems of closed quantum systems reach steady states that correspond to thermal equilibrium.

Advances in the study of optically trapped cold atoms in recent years have made the study of
reasonably well isolated quantum systems experimentally feasible to the extent that ETH has
transformed from a foundational issue in statistical mechanics into a matter of practical
significance. With such an experimental impetus, there is now a sizable literature exploring ETH
mainly using numerical techniques on finite clusters in a variety of many-body local interacting
quantum systems \cite{Rigol_Nature2008, Rigol_PRL2009, BiroliKollathLauchli_PRL10,
  RigolSantos_PRA10, SantosRigol_PRE10, Marquardt_PRE12, Beugeling_scaling_PRE14,
  BrandinoKonikMussardo_PRB12, IkedaUeda_PRE13, SteinigewegPrelovsek_PRE13,
  Beugeling_offdiag_PRE2015, SorgVidmarHeidrichMeisner_PRA14, IkedaHuse_PRL14,
  MondainiSrednickiRigol_PRE16, Alba_PRB15, MondainiRigol_PRA15, lashkari2016eigenstate,
  ArnabSenArnabDas_PRB16, LuitzBarlev_PRL16, d2016quantum, ChandranBurnell_PRB16,
  MondainiRigol_PRE2017}.  There is by now overwhelming evidence that ETH is obeyed over most of the
many-body spectrum in systems with local interactions, with important exceptions close to
integrability and in many-body localized states.

Despite the widespread success of ETH in such numerical investigations, a number of questions remain
open.  In this paper, we extend studies of ETH to a class of many-body interacting models that are
both disordered and nonlocal.  One member of this class is the Sachdev-Ye-Kitaev model, SYK$_4$,
which consists of $N$ Majorana fermions with infinite range random four-point couplings
\cite{Kitaev1,maldacena2016remarks}.  This model recently shot to prominence having been shown to
exhibit a number of remarkable properties \cite{Kitaev1,maldacena2016remarks} including, in the
large $N$ and strong coupling limit (i) an extensive zero temperature entropy (ii) an approximate
emergent reparametrization invariance at large $N$ and (iii) an out-of-time-ordered correlator with
a Lyapunov exponent that saturates a conjectured bound \cite{maldacena2016bound} implying that it is
``maximally chaotic".

A possible intuition for the dynamics of such zero-dimensional disordered models is that the
non-locality allows information to propagate rapidly across all the sites while the randomness in
the couplings translates to dynamics that approximates to the action of random unitaries that should
scramble information rapidly. This intuition is borne out by a numerical study of the quench
dynamics \cite{eberlein2017quantum} which shows that SYK$_4$ approaches equilibrium at some
temperature when the initial state is the equilibrium state for some other model.
However, this intuition fails for the close relative SYK$_2$ which is a random quadratically coupled
model for which the Lyapunov exponent vanishes \cite{2017arXiv170702197G}. The presumption that
scrambling and thermalization should be connected suggests that SYK$_4$ should obey ETH while
SYK$_2$ should not.  This is the question that we explore in this paper. Previous work has showed
that the Majorana version \cite{hunter2017thermalization} and the complex fermion analog of the SYK$_4$ model \cite{sonner2017eigenstate} satisfy ETH. Here, we explore the Majorana version, which splits into different sectors depending on $N$.  We tune parametrically between the  SYK$_4$ and SYK$_2$ models.

Like Refs.~\onlinecite{sonner2017eigenstate,hunter2017thermalization} and the usual ETH literature, we first present a study of the
middle of the many-body spectrum, i.e., excluding the spectral edges.  We carry out a finite size
scaling analysis that tests a strong (scaling) version of ETH \cite{Beugeling_scaling_PRE14}.  Previous
work has argued that the complex fermion analog of SYK$_2$ obeys ETH \cite{Magan_PRL2016}.
Our numerical and analytical work demonstrates scaling behavior of the SYK$_2$ matrix elements that
contrasts sharply with the scaling behavior of ETH but which is very similar to the behavior
observed in other integrable models.  The ETH scaling is valid also at parameters intermediate
between the SYK$_4$ and SYK$_2$ points, which is expected as the model becomes non-integrable as
soon as quartic coupling is included, i.e., everywhere except at the SYK$_2$ point.

In addition to these results pertaining to the \emph{middle} of the spectrum, we show a peculiarity
of the low-energy part of the spectrum, which is more relevant to the holography literature.  The
low-energy eigenstates of SYK$_4$ show the ETH behavior, which is not expected in local models but
may occur in nonlocal models because the low energy eigenstates exhibit volume law entanglement
\cite{Balents_entanglement_arXiv1709}.  The low-energy eigenstates of SYK$_2$ show the scaling
typical of integrable models.  However, at intermediate points, the ETH scaling is no longer seen
--- an admixture of SYK$_2$ destroys ETH scaling at low energies.  We believe this can be traced to
the fact that the SYK$_2$ is an attractive fixed point in the RG sense.  In other words, ETH scaling
of low-energy eigenstates is not governed by non-integrability but by the RG flow.

The paper is organized as follows. In the next section, we outline the statement of ETH,
highlighting the scaling form (size-dependence).  Section~\ref{sec:symm} has a discussion of the
symmetries of SYK$_2$ and SYK$_4$ and the constraints they place on the matrix elements of different
classes of ``local" operators.  Sections \ref{sec:ETH} and \ref{sec:symm} are mostly pedagogical
reviews, meant for readers unfamiliar with ETH scaling or with the finite-size properties of SYK
clusters.  Our main results are in Sections \ref{sec:ondiag}, \ref{sec:highlow} and
\ref{sec:offdiag}.  In Sections \ref{sec:ondiag} and \ref{sec:highlow} we study the scaling
behaviour of the diagonal matrix elements and, in Section~\ref{sec:offdiag}, the off-diagonal matrix
elements. We then conclude with a discussion of some of the significance of our findings.

\section{The Eigenstate Thermalization Hypothesis}
\label{sec:ETH}

ETH is a generic mechanism for thermalization of observables in a closed, finite, non-integrable
system out of equilibrium, i.e., a mechanism for why the long-time average value of an observable
$\hat{O}$ should be describable by a Gibbs ensemble.  The class of operators for which ETH is
expected to be valid includes local operators.  ETH is a statement about the diagonal matrix
elements of the operator in the energy eigenstates $\ket{n}$ of the Hamiltonian, i.e., the
eigenstate expectation values $O_{nn}=\bra{n}\hat{O}\ket{n}$.  ETH states that the diagonal matrix
elements of $O_{nn}$ are smooth functions of the energy in the case when the system size is large,
with the fluctuations being exponentially small in the system size.  This follows from the idea that
high-energy eigenstates of non-integrable systems are complicated enough that the eigenstate
coefficients are effectively random.  For systems with a finite Hilbert space dimension
$\mathcal{D}$, the central limit theorem yields the scaling to be $\mathcal{D}^{-1/2}$
\cite{Marquardt_PRE12, Beugeling_scaling_PRE14}:
\begin{equation} \label{eq:eth_on}
O_{nn} = f^{(1)}_O (E_n) + \frac{1}{\mathcal{D}^{1/2}} f^{(2)}_O (E_n) R_{nn}  
\end{equation}
where $R_{nn}$ is a pseudo-random number with unit width and $f^{(\alpha)}_O(E)$ are smooth functions
of the eigenenergy $E_n$.

A further important ingredient of ETH concerns the off-diagonal matrix elements: $O_{mn}$ for $m\neq
n$.  These matrix elements should be suppressed exponentially in the system size.  Use of the
central limit theorem again shows the scaling to be $\mathcal{D}^{-1/2}$: 
\begin{equation}  \label{eq:eth_off}
  O_{mn} =  \frac{1}{\mathcal{D}^{1/2}} f^{(3)}_O \left(\epsilon, \omega \right) R_{mn}  
\quad \text{for} \; m\neq n 
\end{equation}
where $R_{mn}$ is a pseudo-random number with unit width, $\epsilon=\tfrac{1}{2}(E_m+E_n)$, and
$\omega = E_m-E_n$.  Together, these are usually written together in the following form:
\begin{equation}  \label{eq:eth_usual}
O_{mn} =   f^{(1)}_O (E_n) \delta_{mn} ~+~ e^{-S(E)/2} f_O\left(\epsilon, \omega\right) R_{mn}  \,. 
\end{equation}
This common form is more general than Eqs.\ \eqref{eq:eth_on},  \eqref{eq:eth_off}, because it
applies also to systems with unbounded Hilbert spaces.  For systems with finite Hilbert spaces, the
entropy in the middle of the spectrum scales as $S(E)=\ln\mathcal{D}+\text{const}$, so that the two
forms are equivalent.   

Although this is not commonly stressed, ETH is a scaling statement.  The idea of eigenstate
coefficients being effectively random leads unavoidably to the $\mathcal{D}^{-1/2}$ scaling.  Thus,
a system in which the state-to-state fluctuation of diagonal matrix elements decreases polynomially
with system size would not be considered as satisfying the ETH, as formulated commonly through
Eq.\ \eqref{eq:eth_usual}.  Polynomial scaling has been observed in integrable systems
\cite{ziraldo2013relaxation, Alba_PRB15, ArnabSenArnabDas_PRB16}, which are considered not to obey
the ETH.  Also, diagonal matrix elements being equal to the canonical value in the infinite-size
limit (used as a definition of ETH in Ref.~\cite{Magan_PRL2016}) is not equivalent to the usual
statement of ETH, outlined above.

For local Hamiltonians, the ETH is expected to be valid only in the middle of the spectrum
(high-energy states).  Since we are studying non-local Hamiltonians in this work, we will also
examine whether it is valid for low-energy states.

The physical content of the ETH Ansatz  becomes clear when we consider computing thermodynamic
observables as long time averages. For example, consider $O(t) \equiv \langle \psi(t) \vert \hat{O}
\vert \psi(t) \rangle$ on state $\vert\psi(t)\rangle = \sum_m c_m e^{-iE_m t}\vert m\rangle$: 
\[ 
O(t) = \sum_m \vert c_m \vert^2 O_{mm} + \sum_{\substack{m,n \\ m\neq n}} c_m^\star c_n
e^{i(E_m-E_n)t} O_{mn}.   
\] 
The long-time average of this operator expectation value is
\[ 
\langle O\rangle \equiv \frac{1}{T}\int_0^T dt O(t) = \sum_{m} \vert c_m \vert^2 O_{mm} 
\] 
which is the trace of the operator computed over the so-called diagonal ensemble
\cite{Rigol_Nature2008}. One may show that this is equivalent to the microcanonical average if ETH
is satisfied, assuming that the state has mean energy $E$ with subextensive fluctuations.
The condition of small off-diagonal elements ensures that the time average of $O(t)$ converges to
its thermal value in the long time limit.

ETH in the form stated above has been tested and found to be valid on a range of local many-body
interacting models using, primarily using full spectrum numerical diagonalization on finite
systems. The effect of tuning parameters toward an integrable point has also been widely explored
\cite{Rigol_PRL2009, Rigol_PRA09, RigolSantos_PRA10, Beugeling_scaling_PRE14,
  SorgVidmarHeidrichMeisner_PRA14}.  As the couplings are tuned towards an integrable point, there
is a continuous departure from the conditions of ETH, e.g., a broadening of the distributions of
matrix elements.
The significance of the breakdown of ETH close to integrability is that integrable points have an
extensive number of constants of the motion that prevent thermalization to the usual thermodynamic
ensembles. Instead, such systems are conjectured to thermalize to a generalized Gibbs ensemble
\cite{Rigol_GGE_PRL20017} that enforces the extensive number of integrals of the motion in such
systems. Another class of systems that do not respect ETH are Anderson localized or many-body
localized systems in which disorder inhibits thermalization
\cite{doi:10.1146/annurev-conmatphys-031214-014726}.

In the following, we address the question of whether ETH is respected in a class of models that are
simultaneously fully connected, so that the concept of locality is absent, and intrinsically
disordered.  We test ETH scaling, as described above, for SYK$_2$ and SYK$_4$.


\section{SYK Models at Finite N}
\label{sec:symm}

In this section, we introduce the SYK$_2$ and SYK$_4$ models and their symmetries. We introduce the
few-body operators whose matrix elelments we will study, and show how the model symmetries place
constraints on these matrix elements.

\subsection{Hamiltonian and Symmetries}


In the remainder of this article, we explore the properties of
\begin{equation}
H = (\cos\theta) H_{{\rm SYK}_4} + (\sin\theta) H_{{\rm SYK}_2}
\end{equation}
where 
\begin{equation}
H_{{\rm SYK}_4} = \sum_{1\leq  i<j<k<l \leq N}  J_{ijkl} \chi_i \chi_j \chi_k \chi_l
\end{equation}
and 
\begin{equation}
H_{{\rm SYK}_2} = \sum_{1\leq i<j\leq N} i K_{ij} \chi_i \chi_j.
\end{equation}
Here $\chi_i$'s are Majorana operators.  The couplings $K_{ij}$ and $J_{ijkl}$ are gaussian
distributed random variables with mean zero and respective variances $\langle K^2_{ij} \rangle =
K/\sqrt{N}$ and $\langle J^2_{ijkl}\rangle = J \left( \frac{6}{N^{3}}\right)^{1/2}$.
The parameter $\theta$ runs from $0$ to $\pi/2$, with $\theta=0$ and $\theta=\pi/2$ being the
SYK$_4$ and SYK$_2$ points, respectively.

The anticommutation relation satisfied by Majoranas, $\left\{ \chi_i, \chi_j \right\}=\delta_{ij}$,
is identical to the Euclidean Clifford algebra $\left\{ \Gamma_m, \Gamma_n \right\}=2\delta_{mn}$ up
to normalization.  Therefore, we may represent the $N=2M$ site Majoranas with Hilbert space
dimension $2^{N/2}$ by $N$ gamma matrices of size $2^{N/2}\times 2^{N/2}$. The particular
representation we use for numerical analysis is given below in Section~\ref{sss:repn}.
We introduce the parity operator $P \equiv i^{-N/2} \prod_{i=1}^{N} \Gamma_{i}$ and note (i) that $P^2=1$ (ii) that $P$ anticommutes with the $\Gamma_i$ for a given $N$. Property (ii) implies that the SYK$_4$ Hamiltonian commutes with $P$ so the $\pm1$ parity is a good quantum number.

We now consider the time reversal operator $T=U_{T} K$: an anti-unitary operator which may be
written as the product of a unitary operator $U_T$ and the complex conjugation operator $K$. The gamma matrices obey $K\Gamma_i K = -(-)^i \Gamma_i$. Our
expectation is that complex fermions $c$ map to $c^\dagger$ under time reversal, which implies that
$T\chi_i T^{-1}=\chi_i$.  This is ensured by choosing
\begin{equation} \label{eq_UT1}
U_{T} = P^{M+1} \prod_{m=1} \Gamma_{2m-1}. 
\end{equation}

\subsection{Observables}

Our investigation into the validity of ETH in SYK models rests on calculations of matrix elements for particular few-body operators. The operators we consider are
\[  A_{ij} \equiv i \chi_i \chi_j  \]
and 
\[  B_{ijkl} \equiv  \chi_i \chi_j \chi_k \chi_l  \]
where the Majorana labels should be distinct but are otherwise arbitrary.  

\subsection{Periodicity of SYK$_4$}

\begin{table*}[tbph]
\begin{tabular}{|c||c|c|c|c|c|c|c|c|c|c|c|c|c|c|}
  \hline
$N$ &  4 & 6  & 8 & 10 & 12 & 14 & 16 & 18 & 20 & 22 & 24 & 26 & 28 & 30 \\
   \hline
   Ensemble  & GSE & GUE &  GOE&  GUE&  GSE&  GUE&  GOE&  GUE&  GSE &  GUE&  GOE&  GUE&  GSE & GUE \\   
   \hline
\end{tabular}
\caption{\label{table:ensembletables}
Association of the SYK$_4$ model of different sizes to the three random matrix ensembles.}
\end{table*}

The SYK$_4$ Hamiltonian is a matrix composed of $O(N^4)$ independent random nonzero elements but
with $2^{N/2}\times2^{N/2}$ elements.  The matrices are thus sparse, as is typical for many-body
Hamiltonians.  As is the case with usual condensed matter Hamiltonians, it can still be useful to
compare the symmetries with dense random matrix ensembles (Wigner-Dyson classes).

As pointed out first in Ref.~\cite{YouLudwigXu_PRB17}, the nature of the spectrum and matrix
elements of the SYK$_4$ model have a periodic dependence on $N$, because the underlying symmetries
correspond to different Wigner-Dyson classes (GOE, GUE, GSE), depending periodically on $N$.  The
periodic correspondence is shown in the table on page \pageref{table:ensembletables} for $N$ up to
30.  This correspondence to the random matrix classes has been verified numerically, e.g., through
the level spacing statistics \cite{YouLudwigXu_PRB17, PhysRevD.94.126010, cotler2017black} and from
the spectral form factor $\vert Z(\beta+it) \vert^2$, where $Z$ is the partition function
\cite{cotler2017black}.  Both measures exhibit departures from random matrix predictions above some
Thouless energy scale as is common in many-body models \cite{beenakker1997random}.  In the spectral
form factor, ramp and plateau features occur in both random matrix theory and SYK$_4$ with the
Thouless energy corresponding to short time scales than the ramp onset \cite{cotler2017black}.
Below, we  review the association between the SYK$_4$ model of different sizes to the Wigner-Dyson
classes.  We also present numerical data illustrating features complementary to those in the earlier
literature.

Because the Majoranas are invariant under time reversal, so too is the SYK$_4$ Hamiltonian, $HT=TH$.
By direct calculation one may show that $T^2 = 1$ for $N=0,2$ mod $8$ and $T^2 = -1$ for $N=4,6$ mod
$8$. It also follows that $(PT)^2 =1$ for $N=0,6$ mod $8$ and $(PT)^2 = -1$ for $N=2,4$ mod $8$.  So
there are four distinct cases to consider based on the $T^2$ and $(PT)^2$. We always work in the basis with 
\[
P = \left( \begin{array}{cc} \boldsymbol{1} & \boldsymbol{0} \\ \boldsymbol{0} & -\boldsymbol{1} \end{array} \right).
\]
To see how all the symmetries constrain the spectrum, we now show that the algebraic relations on $T$ and $PT$ may be satisfied by a particular $U_T$ matrix within each class that then fixes the form of the Hamiltonian.

\subsubsection{$N=0$ mod $8$} 

This class is specified by the conditions $T^2=1$ and $(PT)^2=1$. One can find a representation for the gamma matrices for which
\[  
U_T = \left( \begin{array}{cc} \boldsymbol{1} & \boldsymbol{0} \\ \boldsymbol{0} &
    -\boldsymbol{1} \end{array} \right)
\]
which fulfils the algebraic relations on $P$ and $T$.
It follows from time reversal invariance, $U_T H^\star U_T^{-1}=H$, that the Hamiltonian takes the form
\[  
H =  \left( \begin{array}{cc} \boldsymbol{H}_{\mathbb{R}}^{(1)} & \boldsymbol{0} \\ \boldsymbol{0} &
    \boldsymbol{H}^{(2)}_{\mathbb{R}} \end{array} \right). 
\]
Each block is a distinct real random matrix.  Therefore, we expect gross properties of the
eigenstates of each block to correspond to those of the GOE random matrix ensemble.

\subsubsection{$N=2$ mod $8$} In this case, 
\[  U_T = \left( \begin{array}{cc} \boldsymbol{0} & \boldsymbol{1} \\ \boldsymbol{1} & \boldsymbol{0} \end{array} \right)  \]
fulfils the conditions $T^2=1$ and $(PT)^2=-1$ and then $U_T H^\star U_T^{-1}=H$ implies that the Hamiltonian takes the form
\[  H =  \left( \begin{array}{cc} \boldsymbol{H}_{\mathbb{C}} & \boldsymbol{0} \\ \boldsymbol{0} &
    \boldsymbol{H}^{*}_{\mathbb{C}} \end{array} \right). 
\]
The two chirality sectors thus have identical GUE spectra.  In other words, the eigenvalues are doubly
degenerate with distinct chirality eigenvalues. 

\subsubsection{$N=4$ mod $8$}
For this class,
\[  
U_T = \left( \begin{array}{cc} \boldsymbol{\Omega} & \boldsymbol{0} \\ \boldsymbol{0} &
    -\boldsymbol{\Omega} \end{array} \right)  
\]
ensures that $T^2=-1$ and $(PT)^2=-1$ where $\boldsymbol{\Omega}^2 =-1$ which may be, for example, 
\[ \boldsymbol{\Omega} =  \left( \begin{array}{cc} \boldsymbol{0} & \boldsymbol{1} \\ -\boldsymbol{1} & \boldsymbol{0} \end{array} \right). 
\]
Then
\[  
H =  \left( \begin{array}{cc} \boldsymbol{H}^{(1)}_{\mathbb{H}} & \boldsymbol{0} \\ \boldsymbol{0} &
    \boldsymbol{H}^{(2)}_{\mathbb{H}} \end{array} \right). 
\]
where $\boldsymbol{H}^{(1)}_{\mathbb{H}}$ and $\boldsymbol{H}^{(2)}_{\mathbb{H}}$ are each a matrix
of real quaternions represented by $2\times 2$ blocks of complex numbers of the form
\[  
\left( \begin{array}{cc} z_1 & z_2 \\ -z_2^{\star} & z_1^\star \end{array} \right). 
\]
Random Hamiltonians built from real quaternions belong to GSE by definition. The condition $T^2=-1$ enforces a double degeneracy on the eigenstates in a
similar manner to the Kramers degeneracy of time reversal invariant half odd integer spin systems.
The $P=\pm1$ blocks have distinct spectra, but the eigenvalues within each sector are doubly
degenerate.

\subsubsection{$N=6$ mod $8$} 

The fourth and final case has $T^2=-1$ and $(PT)^2=1$ so that
\[  
U_T = \left( \begin{array}{cc} \boldsymbol{0} & \boldsymbol{1} \\ -\boldsymbol{1} &
    \boldsymbol{0} \end{array} \right)  
\]
from which it follows that 
\[  
H =  \left( \begin{array}{cc} \boldsymbol{H}_{\mathbb{C}} & \boldsymbol{0} \\ \boldsymbol{0} &
    \boldsymbol{H}^{*}_{\mathbb{C}} \end{array} \right). 
\]
So, the properties are the same as for the case $N=2$ mod $8$.

\subsubsection{Symmetry constraints on diagonal matrix elements} 

\begin{figure}[tbp]
\begin{center}
\includegraphics[width=\columnwidth]{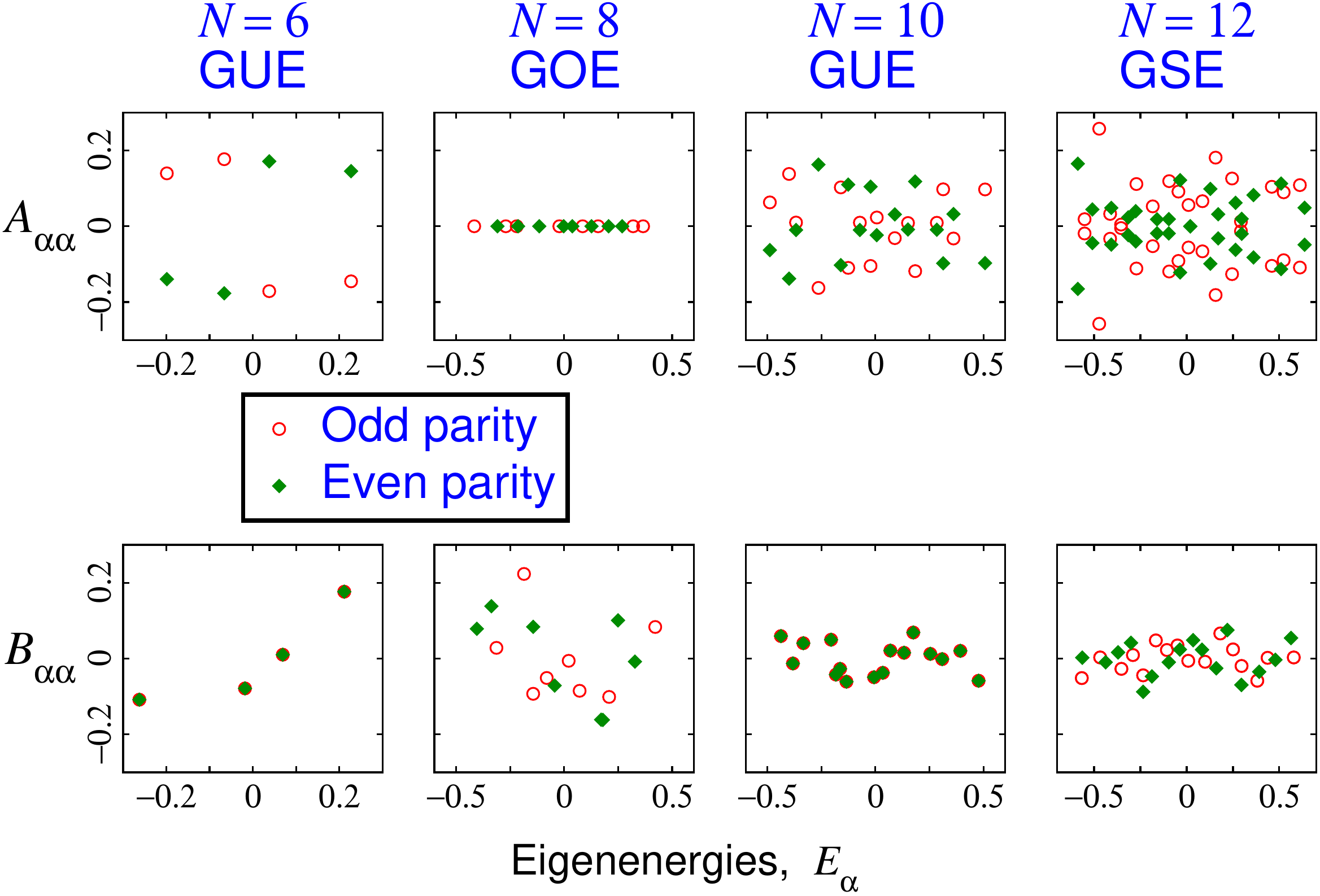}
\caption{ \label{fig:q4:0} Plots of the diagonal matrix elements in the energy basis for SYK$_4$ for small system sizes. These illustrate the appearance of degeneracies within the same sector (GSE) and across different sectors (GUE). The plot also indicates where matrix elements occur in equal or opposite pairs. 
}
\end{center}
\end{figure}

Time reversal places constraints on certain observables. In particular, for the GOE ensemble, since
$T\vert E_n \rangle=\vert E_n \rangle$ and $T i\chi_{i}\chi_{j} T^{-1}= -i\chi_{i}\chi_{j}$, the
diagonal matrix elements vanish, $\langle E_n \vert i\chi_{i}\chi_{j} \vert E_n \rangle = 0$. For
the classes with double degeneracy $\vert+\rangle$, $\vert -\rangle$ ensured by time reversal
symmetry, one finds
\[
\langle + \vert i\chi_{i}\chi_{j} \vert + \rangle = -  \langle - \vert i\chi_{i}\chi_{j} \vert - \rangle
\]
Some of these properties of the SYK$_4$ spectrum and matrix elements are illustrated in
Fig.~\ref{fig:q4:0} and summarized in the table.

Referring to Fig.~\ref{fig:q4:0}, we observe the following features. For $N=8$, the GOE case, energies are singly degenerate and the two-point operator matrix element vanishes as shown above. For the other three system
  sizes, the spectra are doubly degenerate and the two-point matrix elements for the pairs are equal
  and opposite. For the GSE class, the double degeneracy arises for states with identical parities
  as illustrated by the up-down reflection symmetry in the $N=12$ upper panel. For the GUE classes,
  the doubly degenerate states have opposite parities so the upper panels have up-down reflection
  symmetry up to a swap in the symbols (open and closed symbols swap). For the four-point function
  matrix elements for doubly degenerate pairs are the same so for GUE the different parity symbols
  are overlaid and in the GSE case identical parity symbols are overlaid.  

\subsubsection{Numerics on SYK Models}
\label{sss:repn}

We may build a particular gamma matrix representation starting from the Pauli matrices $\sigma_1$, $\sigma_2$ and
$\sigma_3$, the first two of which represent two Majoranas. Then, given the gamma matrices for $N$
Majoranas ${\Gamma^{(N)}_i }$ together with the $P \equiv i^{-N/2} \prod_{i=1}^{N}
\Gamma_{i}$ operator we may obtain a representation for $N+2$ Majoranas
\begin{align*}
\Gamma_i^{(N+2)} & = \sigma_1 \otimes \Gamma_i^{(N)} \hspace{0.5cm} i=1,\ldots, N \\
\Gamma_{N+1}^{(N+2)} & = \sigma_1 \otimes P^{(N)} \\
\Gamma_{N+2}^{(N+2)} & = \sigma_2 \otimes \mathbb{I}.
\end{align*}
In this representation,
\[
P = \left( \begin{array}{cc} \boldsymbol{1} & \boldsymbol{0} \\ \boldsymbol{0} & -\boldsymbol{1} \end{array} \right)
\]
and the Hamiltonian is block diagonal with each block giving states of distinct $P=\pm 1$.


\subsection{SYK$_2$:  Operators and periodicity }

\begin{figure}[tbp]
\begin{center}
\includegraphics[width=\columnwidth]{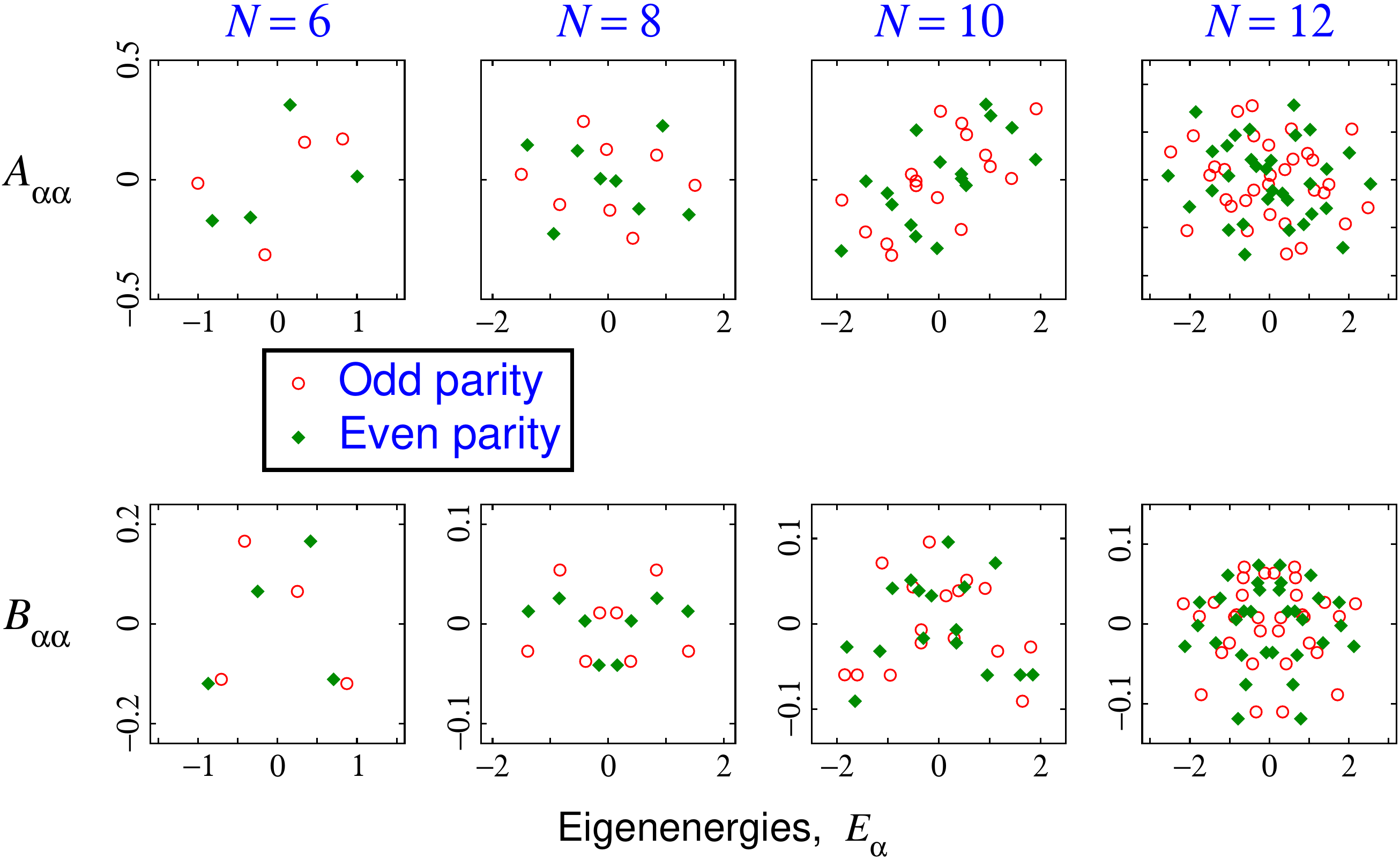}
\caption{Plots of diagonal-in-energy matrix elements for SYK$_2$ for several system sizes.  For each
  case, energies appear in equal and opposite pairs. As demonstrated in the main text, matrix
  elements of the two-point operator (upper row) have a symmetry under $E_{\alpha}\rightarrow
  -E_{\alpha}$ and $\langle A_{\alpha\alpha}\rangle\rightarrow -\langle A_{\alpha\alpha}\rangle$
  while those of a four-point operator (lower row) have a symmetry under $E_{\alpha}\rightarrow
  -E_{\alpha}$ and $\langle B_{\alpha\alpha}\rangle\rightarrow \langle
  B_{\alpha\alpha}\rangle$. Indications of a periodicity in $N$ can be seen in the eigenvalues of
  parity (shown in green and red).} \label{fig:q2:0}
\end{center}
\end{figure}

We now consider the SYK$_2$ model. Once again, the Hamiltonian is block diagonal in the gamma matrix
representation given above because it commutes with $P$. SYK$_2$ is not time reversal invariant but
instead has the feature that $THT^{-1}=-H$ implying that the spectrum has paired eigenvalues with
$\pm E_n$. Despite the lack of time reversal symmetry, the periodicities in $N$ mod $8$ are present
but less distinctive than in SYK$_4$. Similar arguments to those for SYK$_4$ reveal that all the
Hamiltonians have complex entries but $N=0,4$ mod $8$ have $\pm E_n$ pairs arising with the same
parity while $N=2,6$ mod $8$ have $\pm E_n$ pairs coming from different blocks. In SYK$_2$, time
reversal antisymmetry places the following constraints on matrix elements
\begin{align*}
\langle + \vert i\chi_{i}\chi_{j} \vert + \rangle & = -  \langle - \vert i\chi_{i}\chi_{j} \vert - \rangle \\
\langle + \vert \chi_{i}\chi_{j}\chi_{k}\chi_{l}  \vert + \rangle & = \langle - \vert \chi_{i}\chi_{j}\chi_{k}\chi_{l}  \vert - \rangle \\
\end{align*}
where, here, $\vert+\rangle, \vert-\rangle$ have energies with opposite sign and the site indices on
each of the operators are distinct. These features of the matrix elements are illustrated in
Fig.~\ref{fig:q2:0} where $A_{\alpha\alpha}$ denotes some two point Majorana operator and
$B_{\alpha\alpha}$ a four-point Majorana.

The SYK$_2$ model has a Poissonian level spacing distribution as is typical of integrable
models. This is most easily understood in the language of complex fermions. In the single particle
sector, the hopping model is simply a random matrix in GOE with a semi-circle density of states. The
multi-particle sectors are straightforwardly obtained from the single particle sector and the
resulting multiparticle density of states should be gaussian as an application of the central limit
theorem. For a given window of energies and at sufficiently high energies, the occupation number
distribution is effectively random because there is no mutual level repulsion in this integrable
case so the level spacings are Poisson distributed.

\subsection{Combined Hamiltonian (SYK$_4$+SYK$_2$) --- spectrum and level statistics}

\begin{figure}[tbp]
\begin{center}
\includegraphics[width=\columnwidth]{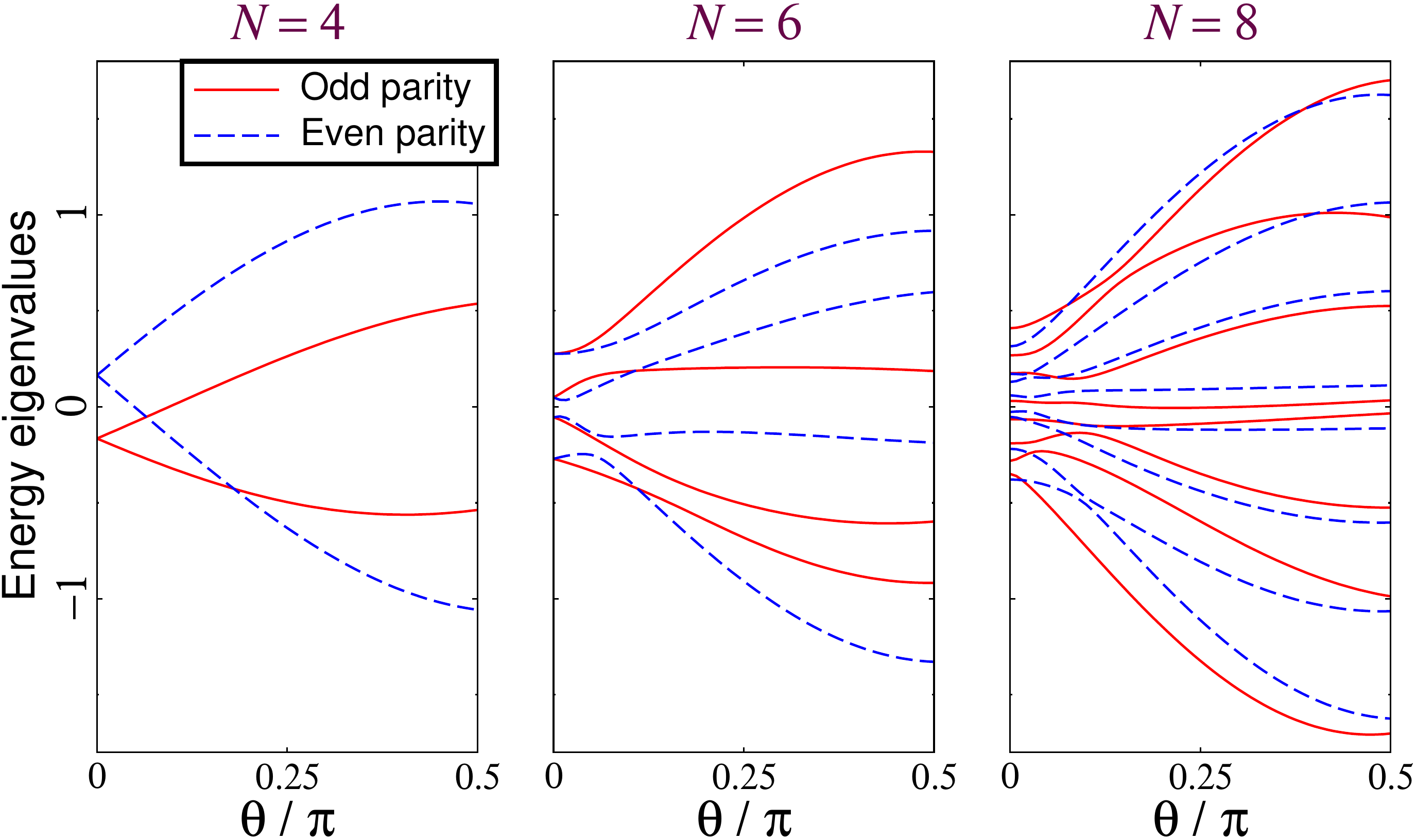}
\caption{Full spectrum of $N=4$, $6$ and $8$ models as $\theta$ is tuned from SYK$_4$ ($\theta=0$)
  to SYK$_2$ ($\theta=\pi/2$) illustrating the presence of symmetries at $\theta=0,\pi/2$, as
  explained in the main text.} \label{fig:q2q4:spectrum}
\end{center}
\end{figure}

Fig.~\ref{fig:q2q4:spectrum} shows the evolution of the spectrum for different small system sizes as
coupling $\theta$ is tuned from $0$ corresponding to SYK$_4$ to $\theta=\pi/2$ which is SYK$_2$. The
parity symmetry is common to models at all $\theta$. For $\theta=0$, the three system sizes
considered correspond to GSE ($N=4$), GUE ($N=6$) and GOE ($N=8$) which respectively have double
degeneracies within each parity sector, double degeneracies within opposite parity sectors and only
accidental degeneracies. As $\theta$ increases, the energies flow to the $\pm E$ symmetry at
$\theta=\pi/2$ which is present in identical parity sectors for $N=4,8$ and opposite parity sectors
for $N=6$. For intermediate $\theta$ there are direct level crossings between different parity
sector eigenstates and avoided level crossings between identical parity sector eigenstates as one
would expect.

\begin{figure}[tbp]
\begin{center}
\includegraphics[width=\columnwidth]{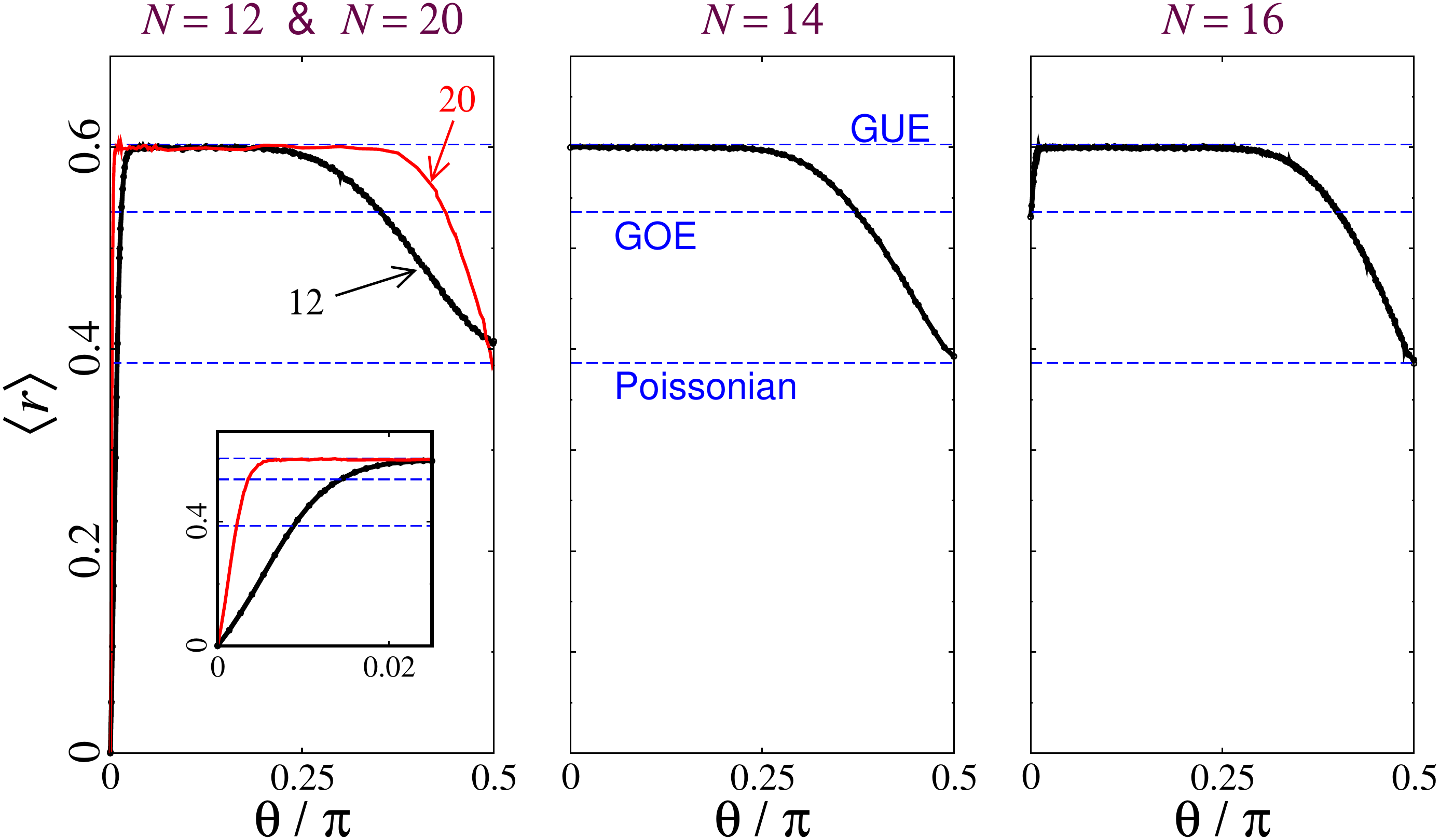}
\caption{Ratio of consecutive level spacings $\langle r\rangle$, as a function of the coupling
  interpolating between SYK$_4$ for $\theta=0$ and SYK$_2$ for $\theta=\pi/2$.  Values for the
  Wigner-Dyson random matrix classes are indicated by dashed horizontal lines.  The three panels
  correspond (from left to right) to GSE, GUE and GOE classes at $\theta=0$.} \label{fig:q2q4:1}
\end{center}
\end{figure}

Further connections between random matrix classes and SYK models are drawn out through a calculation
of the mean ratio of consecutive level spacings $\langle r\rangle$ \cite{PhysRevB.75.155111}. This
measure is based on $s_n = E_{n+1}-E_n$, the set of level spacings in an ordered list ${E_n}$ of
eigenenergies. The ratio
\[  
r_n = \frac{{\rm min}(s_n, s_{n-1})}{{\rm max}(s_n, s_{n-1})}
\]
is defined for each pair of consecutive level spacings. For Poisson statistics, the probability
distribution of $r_n$ is $P(r)=2/(1+r)^2$ with mean $\langle r \rangle=2\ln 2-1 \approx =0.39$. For
the Wigner-Dyson ensembles, we have $\langle r \rangle_{\rm GOE}\approx 0.53$ and $\langle r
\rangle_{\rm GUE}\approx 0.6$ \cite{PhysRevLett.110.084101}.

Fig.~\ref{fig:q2q4:1} shows $\langle r \rangle$ as a function of the coupling $\theta$.  The
spectrum of the odd parity sector is used in each case, and $r_n$ values are combined from a large
number of disorder realizations.
The value of $\langle r\rangle$ corresponds to Poisson statistics at $\theta=\pi/2$ regardless of
Wigner-Dyson class.  At $\theta=0$, the $\langle r \rangle$ value for $N=16$ matches our expectation
that it belongs to GOE. A similar result holds for the GUE cases $N=14$.
For the GSE cases at $\theta=0$, there is a double degeneracy (the Kramers degeneracy) within each
parity sector.  Thus every second level spacing is zero, leading to $\langle r \rangle=0$.

As $\theta$ increases from zero, the statistics in all cases match those of GUE because the matrix
elements are irreducibly complex and the degeneracies are lifted. For larger $\theta$ there is a
smooth crossover to Poissonian statistics.  The Poissonian to GUE crossover near the SYK$_2$ point
has previously also been presented in Ref.~\cite{2017arXiv170702197G}.  The $r$-statistics has also
been described for the SYK$_4$ point in Ref.~\cite{YouLudwigXu_PRB17} and for a coupled chain of
alternating SYK$_4$ and free-Majorana `sites' in Ref.~\cite{HongYao_SYKchain_PRL17}.

Comparing different system sizes (e.g., $N=12$ and $20$ for the GSE class) shows that all the
crossovers become steeper for larger system sizes.  At larger $N$, the deviation from GUE behavior
gets increasingly confined to the singular points $\theta=0$ and $\theta=\pi/2$.


\section{Diagonal Matrix Elements: High-energy states}
\label{sec:ondiag}

In this section we report a study of the eigenstate expectation values
(diagonal matrix elements) in the eigenstates in the middle of the many-body spectrum.  (We refer
to the middle of the spectrum as high-energy states.)  ETH scaling in the low-energy part of the
spectrum is discussed in a later section. 

As usual in studies of the ETH, it is important to restrict to a single symmetry sector.  We present
results for the odd-parity sector.

\subsection{Non-integrable behavior  of SYK$_4$}

\begin{figure}[tbp]
\begin{center}
\includegraphics[width=\columnwidth]{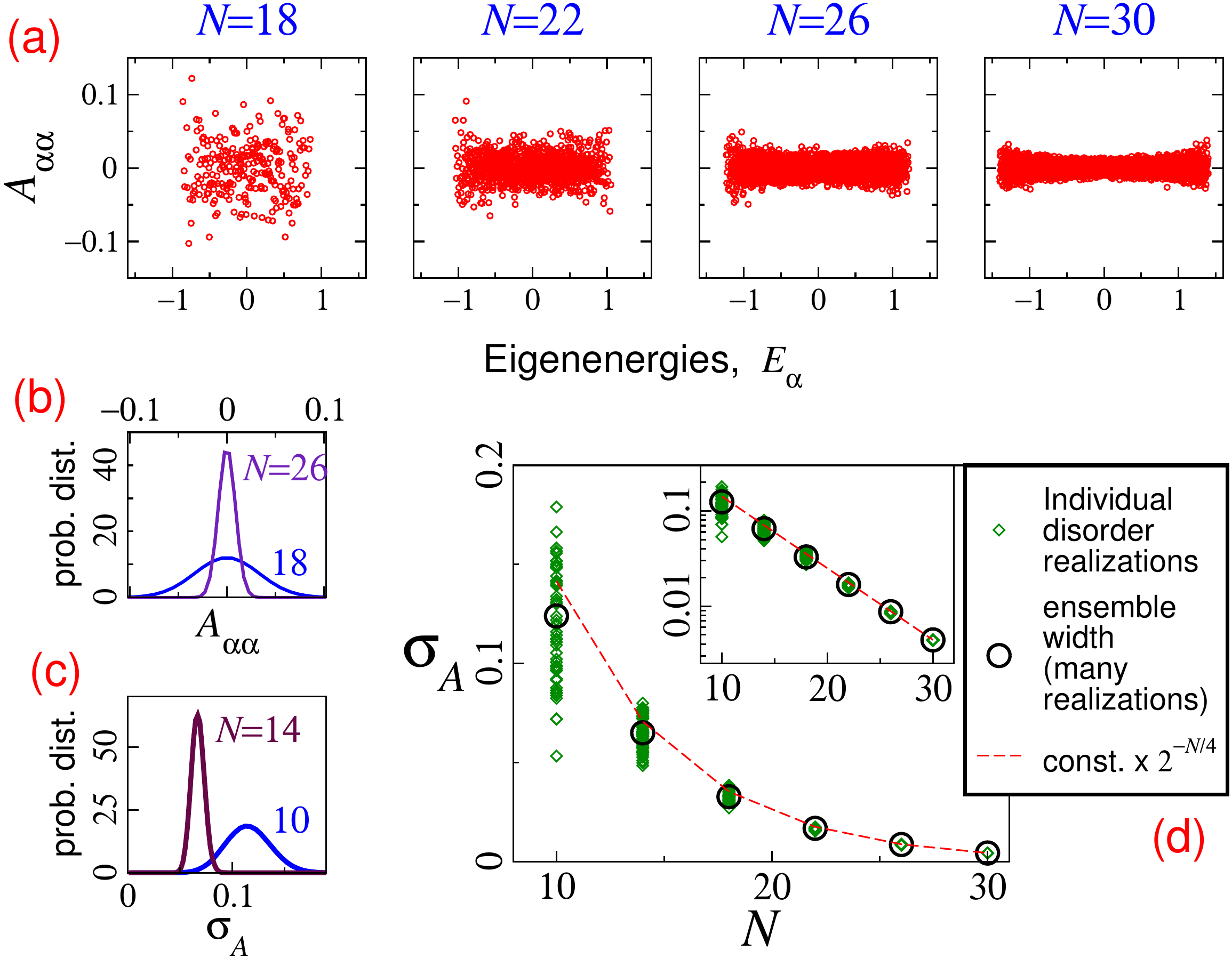}
\caption{\label{fig:q4:1}
(a) Diagonal matrix elements of $A=i\chi_1 \chi_2$ in SYK$_4$ plotted against eigenenergies, for
  system sizes belonging to the GUE ensemble, within a single parity sector for a single disorder
  realization. (b) Distribution of diagonal matrix elements, obtained from combining many disorder
  realizations.
(c) Distribution of $\sigma_A$ values obtained from individual realizations.
(d) Standard deviation $\sigma_A$ of the distribution of diagonal matrix elements of $i\chi_1
  \chi_2$ plotted against system size $N$ for various instances of the GUE ensemble.
Dashed line shows expected ETH scaling for non-integrable systems.
Inset: same data in double-logarithmic scale.
}
\end{center}
\end{figure}

We begin with SYK$_4$.  As we have seen, various properties of SYK$_4$ have a periodic dependence on
the system size $N$ that can be seen to follow from a corresponding random matrix ensemble.  We will
present results for $N$ within a common ensemble.  The GOE systems have the disadvantage that the
$A=i\chi_i\chi_j$ operator is identically zero for all eigenstates, while the GSE systems have
additional symmetry sectors within the odd-parity half of the Hilbert space.  In addition, there are
more numerically accessible finite-size instances of GUE than either of the other two.  Hence we
present results for the systems with GUE symmetry, i.e., for the sequence $N=6, 10, 14, 18,
22. \ldots$. 

We first consider the diagonal matrix elements of the two-point operator $i\chi_i\chi_j$ for some
arbitrary choice of $i$ and $j$. Fig.~\ref{fig:q4:1}(a)  shows the diagonal matrix elements
$A_{\alpha\alpha} ~=~ \langle E_\alpha \vert i\chi_i\chi_j\vert E_\alpha \rangle$ for the full
spectrum from exact diagonalization plotted against energy for four different system sizes falling
into the GUE ensemble. The matrix elements are observed to be distributed around zero for a given
disorder realization independent of the location of the eigenstate within the spectrum. The
distribution has a clear energy dependence, at least for the $N=26$ and $30$, being broader at the
extremes of the spectrum. The distribution of the matrix elements irrespective of energy is gaussian
and the width of the distribution narrows as $N$ increases, as shown in Fig.~\ref{fig:q4:1}(b).

As explained in Section~\ref{sec:ETH}, ETH is a scaling statement --- the width $\sigma_A$ of the
diagonal matrix elements should scale as $1/{\cal D}^{1/2}$.  In the SYK models, ${\cal D}=2^{N/2}$.
Numerical results for six system sizes, shown in Fig.~\ref{fig:q4:1}(d), shows that the width does
indeed scale as $2^{-N/4}$ for SYK$_4$.  (For a single parity sector, the relevant Hilbert space is
half of $2^{N/2}$, i.e., ${\cal D}=2^{N/2-1}$.  However, this is a change of overall factor and does
not affect the scaling form, $2^{-N/4}$.)

There is some shot-to-shot variation of the width $\sigma_A$ as calculated from individual
realizations.  Fig.~\ref{fig:q4:1}(c) shows the distribution of $\sigma_A$ values obtained from
different disorder realizations. The width of these distributions falls off rapidly with system
size.

\subsection{Integrable behavior of SYK$_2$}

\begin{figure}[tbp]
\begin{center}
\includegraphics[width=\columnwidth]{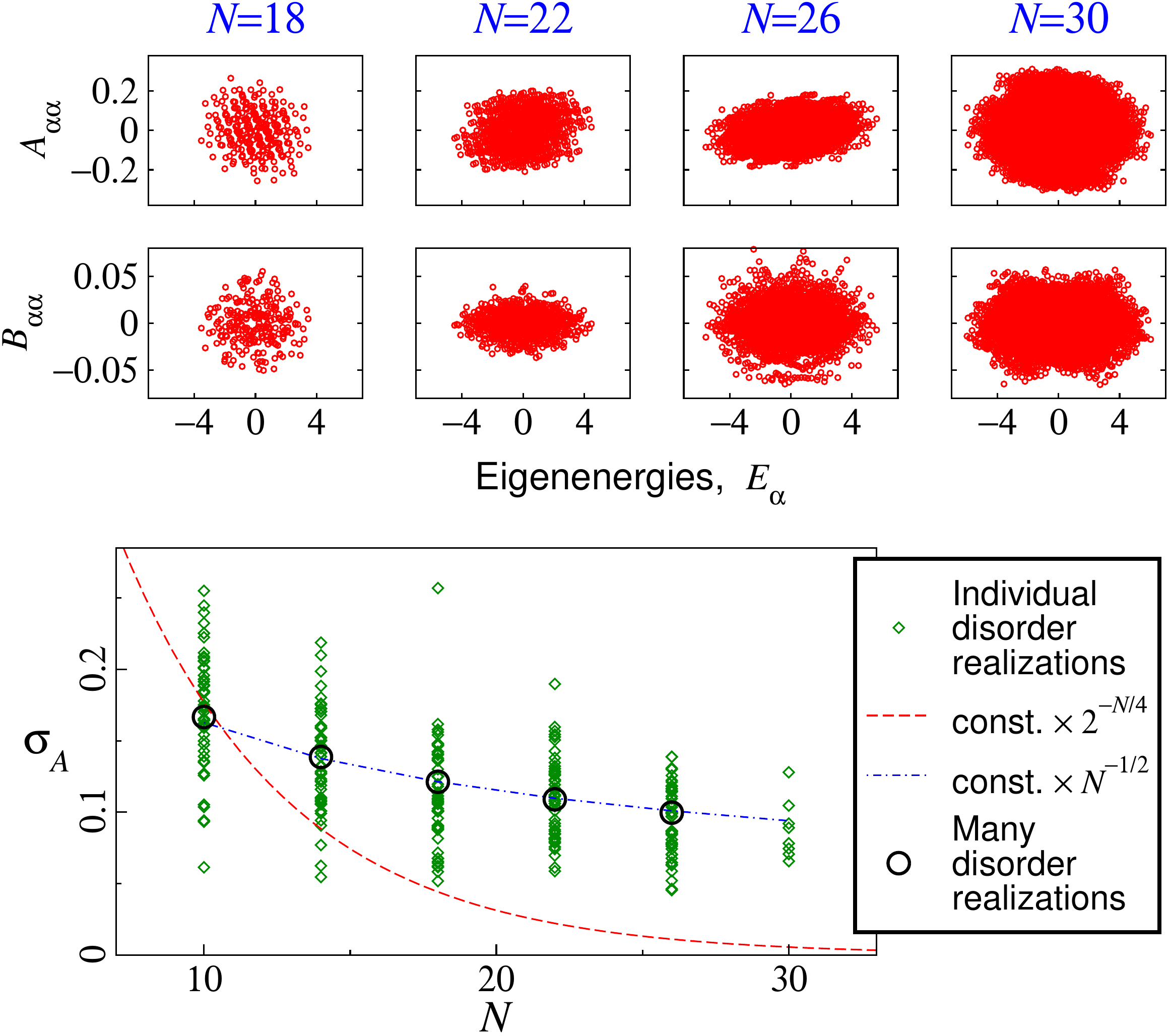}
\caption{ \label{fig:q2:1} Top two rows: Diagonal matrix elements of few-body operators $A$ and $B$
  against corresponding eigenenerges.  Bottom: the width $\sigma_A$ of the distribution of diagonal
  matrix elements plotted against system size.  Two different scaling functions are overlaid for
  comparison.  The size dependence follows $N^{-1/2}$, much slower than the ETH scaling
  $\mathcal{D}^{-1/2}$.}
\end{center}
\end{figure}

We now consider SYK$_2$, which is an integrable (quadratic) model and hence is not expected to
follow ETH scaling.  We consider diagonal matrix elements of the two-point Majorana operator
(operator $A$) and four-point operator (operator $B$).  Numerical diagonalization shows a scatter in
the matrix elements centred on zero (top panels of Fig.~\ref{fig:q2:1}) for both operators and the
distribution does not apparently narrow with increasing system size.  This is qualitatively
consistent with the behaviour observed in other integrable models. A more quantitative picture is
seen by explicitly plotting the width $\sigma_A$ against system size, as in the lower panel of
Fig.~\ref{fig:q2:1}.  This clearly shows that ${\cal D}^{-1/2}$ scaling (i.e., $2^{-N/4}$ scaling)
does not hold.  Different disorder realizations reveal a large scatter in the width for each system
size as shown by the different green points at each $N$. The average of all of these appears to fall
off with a power law $N^{-0.5}$, in strong contrast to standard ETH scaling. For the $B$ operators
we find $N^{-1}$ scaling.

We may derive the scaling of the width of the diagonal matrix element distribution: $1/\sqrt{N}$ for the $A$ operators and $1/N$ for the $B$ operators. We may pair the $N$ Majoranas of SYK$_2$ arbitrarily and here we choose complex fermion $I$ to be composed of Majoranas $i=2I-1$ and $j=2I$ so that $c_{I}=\chi_{2I-1}+i\chi_{2I}$ and $c^\dagger_{I}=\chi_{2I-1}-i\chi_{2I}$. The SYK$_2$ Hamiltonian can be seen to contain number conserving and non-conserving terms. As the Hamiltonian is quadratic we may diagonalize via
\[
c_I = \sum_{\alpha} \left( u_I^\alpha d_\alpha + u_I^{\alpha+N} d^\dagger_\alpha   \right)
\]
so that 
\[
H_{{\rm SYK}_2} = \sum_{\alpha} \epsilon_\alpha d^\dagger_\alpha d_\alpha
\]
and the many-body eigenstates are 
\begin{align*}
  \vert \psi_{\left\{ m \right\}} \rangle = (d_1^\dagger)^{m_1} (d_2^\dagger)^{m_2}
  \ldots (d_{N/2}^\dagger)^{m_{N/2}} \vert 0\rangle  \\  \equiv \vert m_1, m_2,\ldots, m_{N/2}
  \rangle
\end{align*}
corresponding to eigenenergies $\sum_{\alpha} \epsilon_{\alpha} n_\alpha$.
The diagonal matrix element $A_{\alpha\alpha}$ may be chosen without loss of generality to be $\langle\psi_{\left\{ m \right\}} \vert   i\chi_{2I-1}\chi_{2I} \vert \psi_{\left\{ m \right\}} \rangle$ so that the complex fermions belong to the same site
\begin{align*} A_{\alpha\alpha} = \frac{1}{4} \sum_{\alpha} \left\{ \left(  u_I^{\alpha\star} u_I^{\alpha} - u_I^{\alpha+(N/2)} u_I^{\alpha+(N/2) \star}  \right)d^{\dagger}_\alpha d_{\alpha} \right.  \\ \left. + \left(  u_I^{\alpha+(N/2)\star} u_I^{\alpha+(N/2)} - u_I^{\alpha} u_I^{\alpha\star}  \right)d_\alpha d^\dagger_{\alpha}  \right\}
\end{align*}

In the spirit of exploring fluctuations between different eigenstates within the same disorder
realization as is typical in the ETH literature, we make the assumption that the single-particle
eigenstate coefficients $u_{I}^{\alpha}$ are Haar-random variables. Then 
\begin{align*}
& \langle\langle u_I^\alpha u_J^{\beta\star}  \rangle\rangle = \frac{2}{N} \delta_{IJ}\delta^{\alpha\beta} \\
& \langle\langle u_I^\alpha  u_J^{\beta}u_K^\gamma  u_L^{\delta}  \rangle\rangle  = \frac{4}{N^2 -
    4} \left( \delta_{IK}\delta_{JL} \delta^{\alpha\gamma}\delta^{\beta\delta} +
  \delta_{IL}\delta_{JK} \delta^{\alpha\delta}\delta^{\beta\gamma}  \right)  \\ & - \frac{8}{N(N^2 -
    4)}\left( \delta_{IK}\delta_{JL} \delta^{\alpha\delta}\delta^{\beta\gamma}  +
  \delta_{IL}\delta_{JK} \delta^{\alpha\gamma}\delta^{\beta\delta}   \right) 
\end{align*}
where the double brackets $\langle\langle\cdot\rangle\rangle$ indicate an average over the Haar measure. It follows that
$\langle\langle A_{\alpha\alpha} \rangle\rangle=0$ that is, the diagonal two point matrix elements
for SYK$_2$ fluctuate around zero mean. We may compute the variance
$\langle\langle{}A_{\alpha\alpha}^2\rangle\rangle$ of the distribution under the same randomness
assumption with the result that this varies as $1/N$ for large $N$.  This is consistent with our
numerically observed $N^{-1/2}$ scaling for the standard deviation.  A similar calculation for the
four-point operator shows that $\langle\langle{}B_{\alpha\alpha}\rangle\rangle =0$ and
$\langle\langle{}B_{\alpha\alpha}^2\rangle\rangle \sim 1/N^2$. Note that the above results may also
be regarded as coming from a disorder average over the SYK couplings for a fixed energy window.

Taken together, the numerical results we and others have obtained for SYK$_2$ --- the Poissonian
statistics, vanishing of large number of matrix elements and power law fall-off of the spread in the
diagonal matrix elements --- are all consistent with results obtained for integrable many-body
models \cite{d2016quantum,vidmar2016generalized, ziraldo2013relaxation, Alba_PRB15,
  ArnabSenArnabDas_PRB16}.  At integrability, the number of conserved quantities is extensive and
these are thought to bring about thermalization to a generalized Gibbs ensemble, for a wide range of
observables and initial conditions, coinciding with a failure of ETH scaling. We note that
Ref.~\onlinecite{Magan_PRL2016}, which claims ETH in complex SYK$_2$, is based on a different and
non-standard statement of ETH as discussed in Section~\ref{sec:ETH}. We discuss thermalization at
integrable points further in the concluding section \ref{sec_discussion}.

\subsection{Between the SYK$_4$ and SYK$_2$  points}

Our general Hamiltonian, $H = (\cos\theta) H_{{\rm SYK}_4} + (\sin\theta) H_{{\rm SYK}_2}$, is not
integrable, and therefore the middle of the spectrum should show standard ETH scaling and generic
thermalization behavior.  The $\sigma_A\sim\mathcal{D}^{-1/2}$ behavior is demonstrated in Figure
\ref{fig:lowhigh}, top center.  For large enough systems one expects this to be true for all values
of $\theta\in[0,\pi/2)$, i.e., excepting the integrable point $\theta=\pi/2$.  At points closer to
  $\theta=\pi/2$, one expects the $\mathcal{D}^{-1/2}$ behavior to set in at larger sizes.  This is
  the typical behavior close to integrability, as explored, e.g., in
  Ref.~\cite{Beugeling_scaling_PRE14}.

\section{Mid-spectrum versus low-energy matrix elements}
\label{sec:highlow}

\begin{figure*}[btph]
\begin{center}
\includegraphics[width=0.8\textwidth]{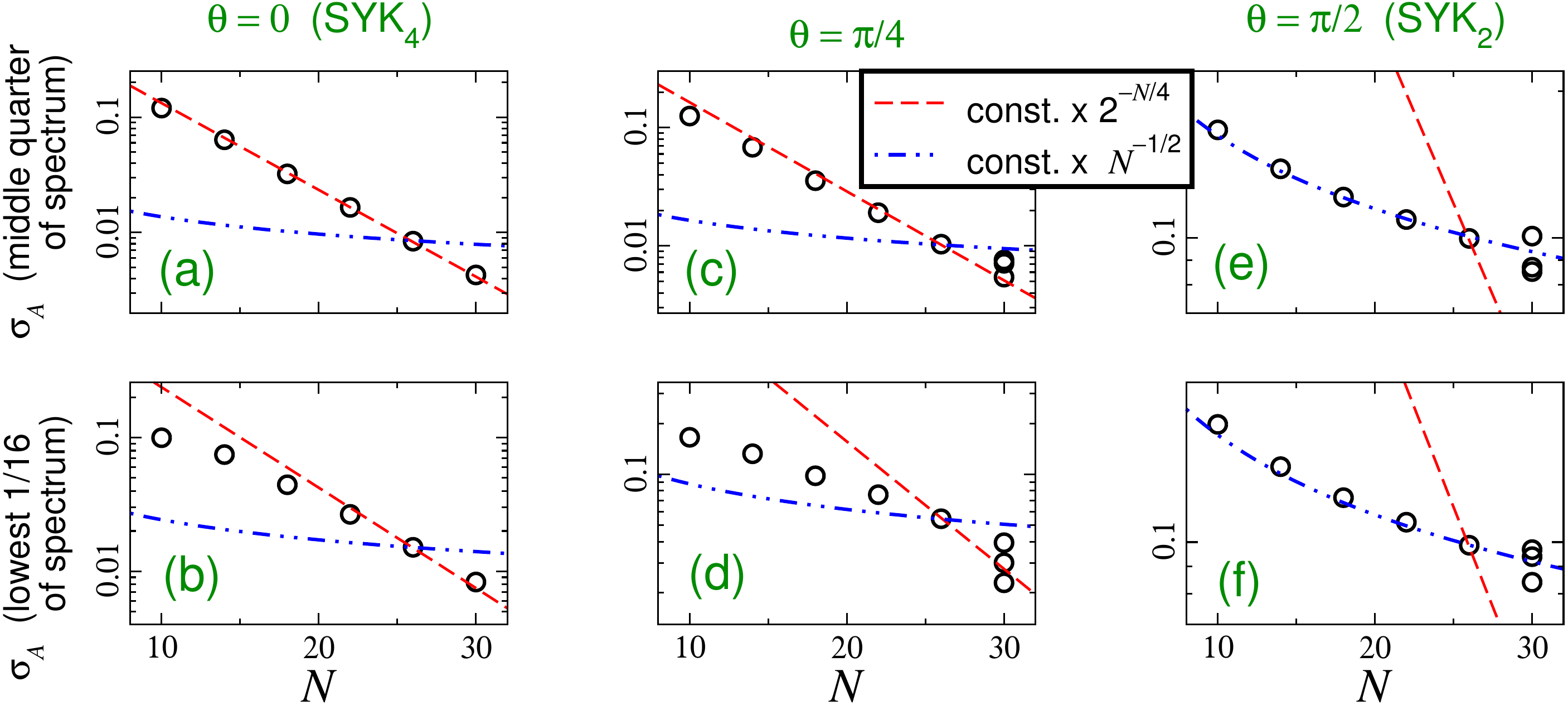}
\caption{\label{fig:lowhigh}
Standard deviation of the distribution of matrix elements of operator $A$, against system size
$N$. Upper panels: middle quarter of the spectrum (``high-energy'' states).  Lower panels: bottom
$1/16$th of the spectrum (``low-energy'' states).
The three columns correspond to $\theta=0$ (SYK$_4$), $\theta=\pi/4$ (intermediate) and
$\theta=\pi/2$ (SYK$_2$).
}
\end{center}
\end{figure*}

Fig.~\ref{fig:lowhigh} shows the scaling of the standard deviation $\sigma_A$ of the
distribution formed from the $A$ operator matrix elements at $\theta=0, \pi/4$ and $\pi/2$ points. The
upper and lower panels show, respectively, $\sigma_A$ drawn from the middle quarter and lower
$1/16$th of the full spectrum. 

Unlike Hamiltonians with local interactions, the SYK$_q$ models (with $q=4,6,8,...$) have the
property that low-energy states in the large-$N$ limit are `thermal' and even maximally chaotic
\cite{Kitaev1,maldacena2016remarks,kourkoulou2017pure}.  In finite-size SYK systems, this is manifested, for example, in the ground-state
entanglement entropy between two halves of the system growing linearly with system size, i.e.,
`volume law'-like behavior instead of `area law'-like behavior \cite{Balents_entanglement_arXiv1709}.
It is therefore natural to conjecture that the low-energy eigenstates might display ETH scaling,
i.e., that the low-energy part of the spectrum might show behavior typical of effectively random
states.  Fig.~\ref{fig:lowhigh}(b) shows ETH scaling for the low-energy eigenstates of SYK$_4$.
While the asymptotic behavior is apparently $\mathcal{D}^{-1/2}$, the behavior sets in at larger
sizes than for the middle of the spectrum.  For local Hamiltonians, the low energy states show quite
different behavior from typical states in the middle of the spectrum - the low energy states tend to
have area law entanglement entropy and ETH ceases to hold. The SYK$_4$ model is exceptional in this
sense since the ETH scaling is present throughout the spectrum.

For $\theta=\pi/2$ (SYK$_2$), $\sigma_A$ shows the previously reported $1/\sqrt{N}$ scaling, not
only at the middle of the spectrum as reported in Fig.~\ref{fig:q2:1}, but also at low energies,
Fig.~\ref{fig:lowhigh}(e,f).

The most interesting behavior is seen for the combined Hamiltonian, as exemplified by the
$\theta=\pi/4$ data in Fig.~\ref{fig:lowhigh}(c,d).  At the $\theta=\pi/4$ point, in the middle of
the spectrum both operators exhibit the $2^{-N/4}$ scaling we have come to expect from the SYK$_4$
model; indeed $1/\sqrt{{\cal D}}$ is the expected behavior for any non-integrable system.  However,
the low-energy behavior is determined not by the non-integrability but rather by the RG properties.
The scaling dimension of the fermion operator in SYK$_4$ is $\Delta=1/4$, so the quadratic SYK$_2$
perturbation has positive mass dimension and is RG relevant. This implies that low energy
observables will behave differently to expectations coming from SYK$_4$ when the SYK$_2$ coupling is
present.  Indeed, Fig.~\ref{fig:lowhigh}(d) shows that, when we consider only states from the lowest
$1/16$th of the spectrum, the scaling of the widths departs considerably from the SYK$_4$ prediction
$\mathcal{D}^{-1/2}$, consistent with the expectation from the RG argument above.  Thus, for
$0<\theta<\pi/2$, the high-energy physics is governed by non-integrability, while the low-energy
physics is governed by flow to the integrable fixed point.

\section{Off-Diagonal Matrix Elements}
\label{sec:offdiag}

\begin{figure}[tbp]
\begin{center}
\includegraphics[width=\columnwidth]{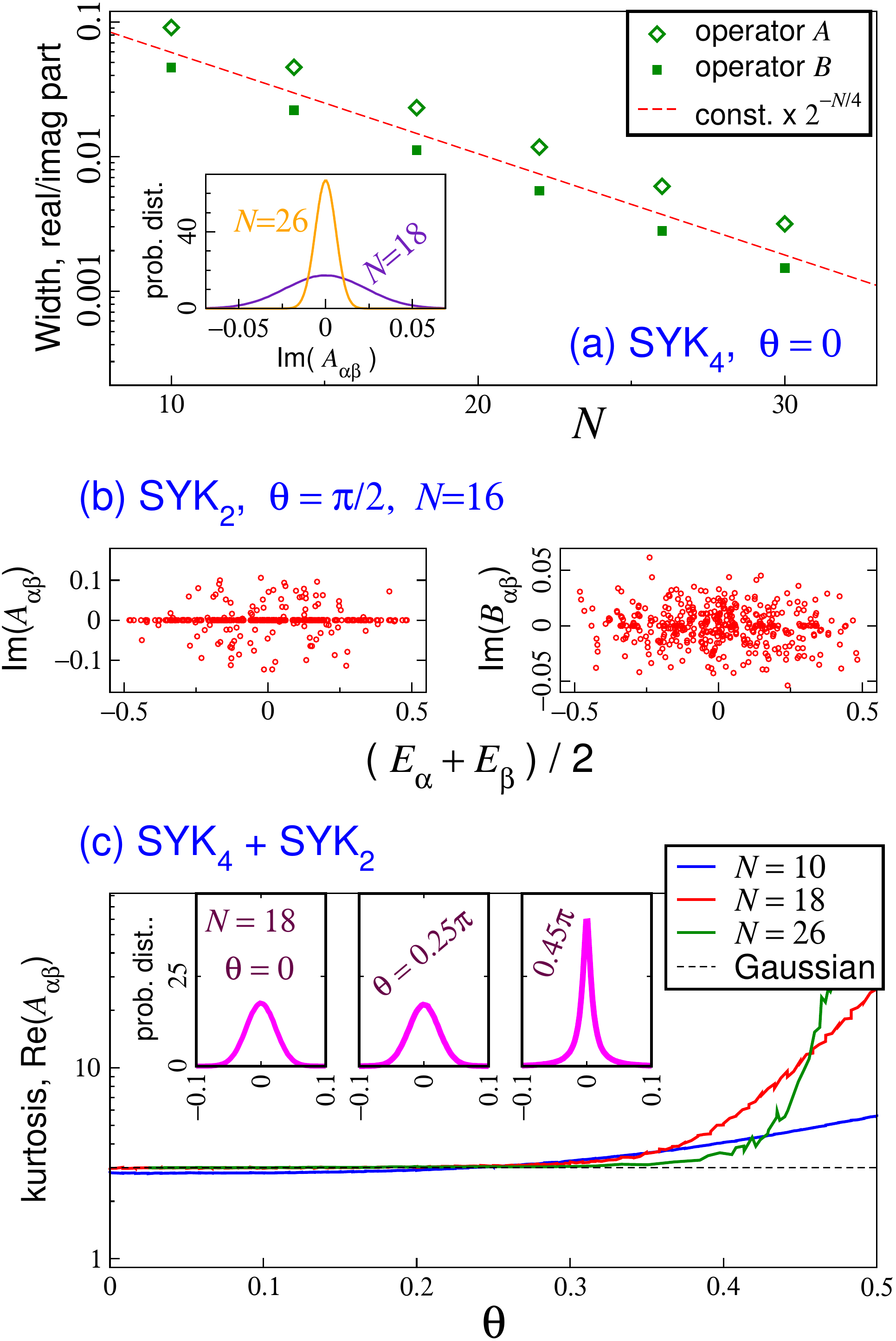}
\caption{ \label{fig:q4:3}
(a) Size-dependence of the width of the distribution of off-diagonal matrix elements of operators
  $A$ and $B$, for the SYK$_4$ model with $N$ values within the GUE ensemble.  Eigenstates in the
  middle quarter of the spectrum are used.  The distribution (inset) is gaussian.
(b) Scatter plot of off-diagonal matrix elements of operators for SYK$_2$. Especially in the
  left-hand panel one can see the large number of vanishing elements, explained in the main text.
(c) The kurtosis of the distribution of off-diagonal matrix elements as a function of $\theta$,
  showing strong non-gaussianity in the neighborhood of the SYK$_2$ model.  The insets show the
  distribution for three values of $\theta$, for $N=18$.
}
\end{center}
\end{figure}

In this section, we discuss the off-diagonal matrix elements, Fig.~\ref{fig:q4:3}, for the operators
$A$ and $B$. 
The off-diagonal matrix elements have arbitrary phase, since the eigenstates each are defined
only up to a phase.  On average, this guarantees that the real and imaginary parts of the
off-diagonal elements will be distributed symmetrically around zero.  

For SYK$_4$, the distribution of off-diagonal elements is gaussian, as in other non-integrable
models \cite{MondainiRigol_PRE2017, Beugeling_offdiag_PRE2015}.  As for the case of diagonal
matrix elements, we study the finite-size scaling of the standard deviation of this
distribution. Once again, we find $\mathcal{D}^{-1/2}$ or $2^{-N/4}$ scaling,
Fig.~\ref{fig:q4:3}(a), consistent with ETH scaling.  The behaviour of the two operators differ only
in the coefficient.

For SYK$_2$, one finds that most off-diagonal matrix elements vanish, Fig.~\ref{fig:q4:3}(b).  Since
SYK$_2$ is a free fermion model, the eigenstates may be organized by quasiparticle number. The $A$
operator written in the quasiparticle basis has number conserving terms and terms that change the
quasiparticle number by two. For the $B$ operator, there are terms that change the quasiparticle
number by zero, two and four. Therefore off-diagonal matrix elements in $A$ may be non-vanishing
only when the eigenstates differ by two in the quasiparticle number which is an exponentially small
number of possible off-diagonal matrix elements. Similar statements hold for any $q$-Majorana
operator for $q\ll N$.

Because of this anomalously large weight at zero, the distribution of off-diagonal matrix elements
has a delta function at zero, and so cannot be meaningfully plotted from finite-size data.  However,
for $\theta$ close to $\pi/2$ (close to the SYK$_2$ point), the distribution retains signatures of
this feature, as seen in the $\theta=0.45\pi$ inset to Fig.~\ref{fig:q4:3}(c).  This is quantified
by plotting the kurtosis of the distribution, which is 3 for a Gaussian distribution, as a function
of $\theta$, Fig.~\ref{fig:q4:3}(c).  (With the number of disorder realizations used, the kurtosis
curves are not completely smooth, but the overall behavior is clear.)  The highly peaked structure
due to the many zeros at/near the SYK$_2$ point leads to a sharp rise of the kurtosis.  Comparison
of the three sizes also shows that the region of proximity shrinks with increasing system size $N$
--- for larger system sizes, a given departure from gaussian occurs for larger $\theta$.

A similar feature was observed near integrability in Ref.~\cite{Beugeling_offdiag_PRE2015} for
multiple non-random local models.  This suggests that our explanation above for zero off-diagonal
elements can be adapted to a wide class of integrable models.

\section{Discussion}
\label{sec_discussion}

In this paper, we have extended tests of ETH to a class of random zero dimensional interacting models that may be tuned from chaotic to integrable. Our detailed numerical analysis shows that the chaotic model SYK$_4$ obeys the standard scaling form of the eigenstate thermalization hypothesis. The key results are that the two and four point matrix elements in the eigenstate basis scale as $2^{-N/4}$ leaving a sharp distribution around zero mean for large finite $N$ both at low energies and around the middle of the spectrum. In contrast, the scaling behaviour of the integrable SYK$_2$ model matrix elements is algebraic in $N$ over the entire spectrum. 

The fact that SYK$_4$ obeys ETH indicates that the model has thermal correlators in the long time
limit even within single eigenstates. SYK$_2$, in contrast, might be expected to thermalize to a
generalized Gibbs ensemble averages determined by the extensive number of conserved quantities. It
turns out that the canonical and GGE predictions coincide in free fermion models for single particle
operators \cite{ziraldo2013relaxation}, such as our $A$ operator and the operator considered in
Ref.~\onlinecite{Magan_PRL2016}. It may be interesting to explore further the quench dynamics of
non-quadratic operators in complex and Majorana SYK$_2$.

The off-diagonal matrix elements in the energy basis determine the way in which subsystems approach
equilibrium. Since ETH is obeyed by SYK$_4$ one may make statements about the long-time behavior of
different correlation functions for example, as observed in Ref.~\onlinecite{cotler2017black} that
the time-dependent two-point correlators tend to zero in the long time limit for $N$ (mod $8) =
0,4,6$ while a nonvanishing limit is possible for the $N$ (mod $8) = 2$ sequence. 

The off-diagonal matrix elements also control the behaviour of thermal out-of-time-ordered
correlation functions (OTOCs) such as $\langle [ A(0), B(t)]^2 \rangle_\beta$ at inverse temperature
$\beta$. Such correlation functions have been much studied recently in the context of quantum chaos
and scrambling. At early times, the local operators mutually commute while the unitary evolution
causes a spreading of information in sufficiently chaotic systems leading to an exponential increase
in the OTOC - the exponent having the interpretation of a Lyapunov exponent $\lambda$ in the
semiclassical limit. One might expect such scrambling of information to be an integral part of
thermalization.  The fact that the Lyapunov exponent is finite in SYK$_4$ and zero in SYK$_2$
\cite{2017arXiv170702197G} is intuitively consistent with these expectations.

SYK$_4$ has the remarkable feature in the large $N$ and $\beta J$ limits that the Lyapunov exponent corresponds to maximal chaos $\lambda=2\pi/\beta$. In contrast, it has been observed in numerical computations of the OTOC that the finite $N$ behaviour of SYK$_4$ has a Lyapunov exponent that is only weakly temperature dependent. The gulf between the latter result and the large $N$ limit cannot be bridged through the finite size scaling admitted by exact diagonalization. 

Similarly, there is only a weak sense in which SYK$_4$ for small system sizes is a particularly
efficient thermalizer. In local many-body interacting systems, insofar as ETH has been tested, one
finds similar behavior to SYK$_4$ in non-integrable models while SYK$_2$ behaves much like local
integrable models. However, significant departures are expected at very low energies between the
random $0+1$D models considered here and local Hamiltonians. This is because the low energy
eigenstates of SYK$_4$ exhibit a volume law entanglement entropy \cite{fu2016numerical,
  Balents_entanglement_arXiv1709} whereas local models have, instead, area law entanglement at low
energies. By showing that ETH scaling holds down to the bottom of the spectrum we have uncovered a new aspect of this unusual low energy behavior. One might speculate that these features are present in generic fully connected and nonintegrable interacting systems.

We have explored ETH for parameters interpolating between SYK$_4$ and SYK$_2$. Remarkably, the low energy behavior departs from the usual ETH scaling because of the RG relevance of SYK$_2$ whereas the middle of the spectrum of the mixed Hamiltonian is governed by nonintegrability and ETH scaling persists.

\begin{acknowledgments}
  We acknowledge conversations with A.~B\"{a}cker, J.~Bardarson, W.~Beugeling, S.~Bhattacharjee, R.~Dantas, M.~Heyl,
  I.~Khaymovich, A.~Lazarides, I.~Mandal, R.~Moessner, P.~Surowka, and D.~Trapin.
\end{acknowledgments}

\bibliography{references}

\end{document}